\documentclass[a4paper, twocolumn]{article}
\usepackage[margin=2cm]{geometry}

\usepackage[style=numeric-comp, url=false, doi=true, isbn=false, sorting=none, maxcitenames=2, maxbibnames=30, giveninits=true]{biblatex}
\renewbibmacro{in:}{}
\DeclareFieldFormat{pages}{#1}
\usepackage{setspace}
\usepackage{amsmath}
\usepackage[version=4]{mhchem}
\usepackage{siunitx}
\usepackage{dirtytalk}
\usepackage[switch]{lineno}
\usepackage{authblk}
\usepackage{graphicx, wrapfig, rotating}
\usepackage[font=footnotesize]{caption}
\usepackage{float}
\usepackage{booktabs}
\usepackage{enumitem}
\setlist{nosep}
\usepackage{abstract}
\usepackage[colorlinks=true, citecolor=green, linkcolor=black, urlcolor=blue]{hyperref}

\usepackage{tikz}
\usetikzlibrary{shapes.geometric, arrows}
\usepackage{bm}
\usepackage{esvect}

\providecommand{\keywords}[1]{\textbf{Keywords:} #1}

\begin{document}

\pagenumbering{arabic}
\date{}
\title{A sensitivity analysis of the PAWN sensitivity index}
\author[1,2]{Arnald Puy\thanks{Corresponding author}}
\author[3, 4]{Samuele Lo Piano}
\author[2,4]{Andrea Saltelli}

\affil[1]{\footnotesize{\textit{Department of Ecology and Evolutionary Biology, M31 Guyot Hall, Princeton University, New Jersey 08544, USA. E-Mail: apuy@princeton.edu}}}

\affil[2]{\footnotesize{\textit{Centre for the Study of the Sciences and the Humanities (SVT), University of Bergen, Parkveien 9, PB 7805, 5020 Bergen, Norway.}}}

\affil[3]{\footnotesize{\textit{Open Evidence, Universitat Oberta de Catalunya, Edifici 22@, Universitat Oberta de Catalunya, 08018 Barcelona, Spain.}}}

\affil[4]{\footnotesize{\textit{School of the Built Environment, University of Reading, JJ Thomson Building, Whiteknights Campus, Reading, RG6 6AF, UK.}}}

\twocolumn[
  \begin{@twocolumnfalse}
\maketitle
\begin{abstract}

The PAWN index is gaining traction among the modelling community as a sensitivity measure. However, the robustness to its design parameters has not yet been scrutinized: the size ($N$) and sampling ($\varepsilon$) of the model output, the number of conditioning intervals ($n$) or the summary statistic ($\theta$). Here we fill this gap by running a sensitivity analysis of a PAWN-based sensitivity analysis. We compare the results with the design uncertainties of the Sobol' total-order index ($S_{Ti}^*$). Unlike in $S_{Ti}^*$, the design uncertainties in PAWN create non-negligible chances of producing biased results when ranking or screening inputs. The dependence of PAWN upon ($N,n,\varepsilon, \theta$) is difficult to tame, as these parameters interact with one another. Even in an ideal setting in which the optimum choice for ($N,n,\varepsilon, \theta$) is known in advance, PAWN might not allow to distinguish an influential, non-additive model input from a truly non-influential model input.

\end{abstract}
\vspace{0.3cm}
\vspace{0.5cm}
\end{@twocolumnfalse}
  ]
  
\saythanks
  
\section{Introduction}
\textcite{Pianosi2015, Pianosi2018} have recently published in \textit{Environmental Modelling \& Software} a new measure for sensitivity analysis, the PAWN index. Like other moment-independent approaches (i.e. entropy-based \parencite{Liu2006a}, density-based \parencite{Borgonovo2014, Borgonovo2007a}), PAWN does not resort to statistical second-order moments such as variance to apportion output uncertainty to the model parameters. Instead, it relies on Cumulative Distribution Functions (CDFs) to characterize the maximum distance between the unconditional output distribution $Y_U$, i.e. obtained by moving all parameters simultaneously, and the conditional output distribution $Y_{C_{ij}}$, i.e. obtained by fixing the $i$-th parameter to $j=1,2,...,n$ values or intervals within its uncertainty range. The difference between $Y_U$ and $Y_{C_{ij}}$ is assessed via the Kolmogorov-Smirnov test, although other distance-based tests, such as the Anderson-Darling's, may also be used~\parencite{KhorashadiZadeh2017}. The final PAWN index for a given parameter is obtained by calculating the mean, the median, the maximum or any other summary statistic over all the KS values computed between $Y_U$ and $Y_{C_{ij}}$.

The most up-to-date approximation to the PAWN index, named \say{the generic approach}~\parencite{Pianosi2018}, is as follows: let there be an $\textbf{A}$ matrix with $v=1,2,...,N$ rows and $i=1,2,...,k$ parameters. After computing the model output $Y$, the range of variation of the $i$-th parameter is split into $j=1,2,...,n$ intervals of size $N_c$ (where $N_c \approx N/n$). The model output linked to the $j$-th interval is used as the conditional model output $Y_{C_{ij}}$. The unconditional model output $Y_U$ can concur with the whole model output or can be a random sub-sample of the same size as $N_c$. With this approach, the total number of model runs to compute PAWN is fully determined by $N$ \parencite{Pianosi2018}.

Based on trials with the~\textcite{Liu2006a} function, the~\textcite{Ishigami1990} function, the SWAT model~\parencite{KhorashadiZadeh2017} or a wind-energy converter model~\parencite{Holl2016}, it has been observed that PAWN might reach convergence much faster than Sobol' indices. A key question, however, is to know how the selection of $N,n$, the sampling of $Y_U$ or the summary statistic affects the accuracy of PAWN.  This is timely given the widespread adoption of the index: since its inception in 2015, PAWN has been cited 94 times, with the number of citations stably increasing from 5 in 2015 to 31 in 2019. Most of the works quoting PAWN are from the environmental sciences (53), followed by engineering (24) and computer science (23) (Scopus search on September 25 2019). Gaining a systematic insight into the internal functioning of PAWN shall thus allow the modelling community to better appraise its robustness, thus increasing its transparency as well as our awareness of its advantages and limitations.

Here we assess the sensitivity of PAWN to the main structural uncertainties involving its calculation, an exercise that might be termed \say{a sensitivity analysis of a sensitivity analysis} (SA of SA). This expression was used by \textcite{Paleari2016} to study how sensitivity indices are affected by uncertainties in the probability distributions used to describe the model inputs, a work that actually falls into the tradition of \say{probability of probability} or \say{probability of frequency} described in  \textcite{Kaplan1981} (see also \textcite{Aven2020}). A similar analysis of the sensitivity of results to changes in the range of input factors can be found in \textcite{Shin2013}. Other approaches include exploring several sensitivity measures \cite{Saltelli1992, Saltelli1993, Pappenberger2008}. More recently, \textcite{Noacco2019}  included considerations of form of the output, sample size, choice of method, measuring interactions, range of distributions, inclusion or exclusion of non-behavioural runs (simulations), and use of dummy variables. 

Our experiment departs from previous studies in that it assesses the uncertainties embedded in the structural design of sensitivity indices, PAWN in that case. We thus explore the implications of the last ring in the chain of uncertainties characterizing any sensitivity analysis, the last stretches of the garden of forking paths that need to be crossed to arrive at a result in any modelling exercise \cite{Borges1941, Gelman2013}. Exploring all sources of uncertainty is indeed an important pursuit, including the notions of context, purpose and motivations as suggested in sensitivity auditing \cite{Saltelli2013}. However, we believe that the papers just reviewed \cite{Paleari2016, Pappenberger2008, Shin2013, Noacco2019} are not a SA of a SA but more general instances of uncertainty exploration. We take the meaning of SA of SA literally (i.e. the analysis of the sensitivity of a sensitivity analysis to its own design parameters), and thus exclude in this paper the exploration of sensitivities to changing methods, assumptions or model designs, however worthy these analyses might be.

\section{Materials and methods}

In order to cross-check our approach, we match the design uncertainties of PAWN against the design uncertainties of the Sobol' total-order index ($S_{Ti}^*$), a well-established measure to determine how much a given model parameter interacts with the rest\footnote{The asterisk in $S_{Ti}^*$ is used to differentiate between Sobol' total-order index ($S_{Ti}^*$), i.e. the index against which PAWN is matched, and the index used to assess the extent to which $S_{Ti}^*$ interacts with its design parameters ($N,\theta$), $S_{Ti}$.}. Although different in scope and nature, the fact that unconditional CDFs are also affected by interactions paves the way for the uncertainties in PAWN and $S_{Ti}^*$  to be explored in parallel. 

\begin{figure*}[ht]
\centering
\includegraphics[keepaspectratio]{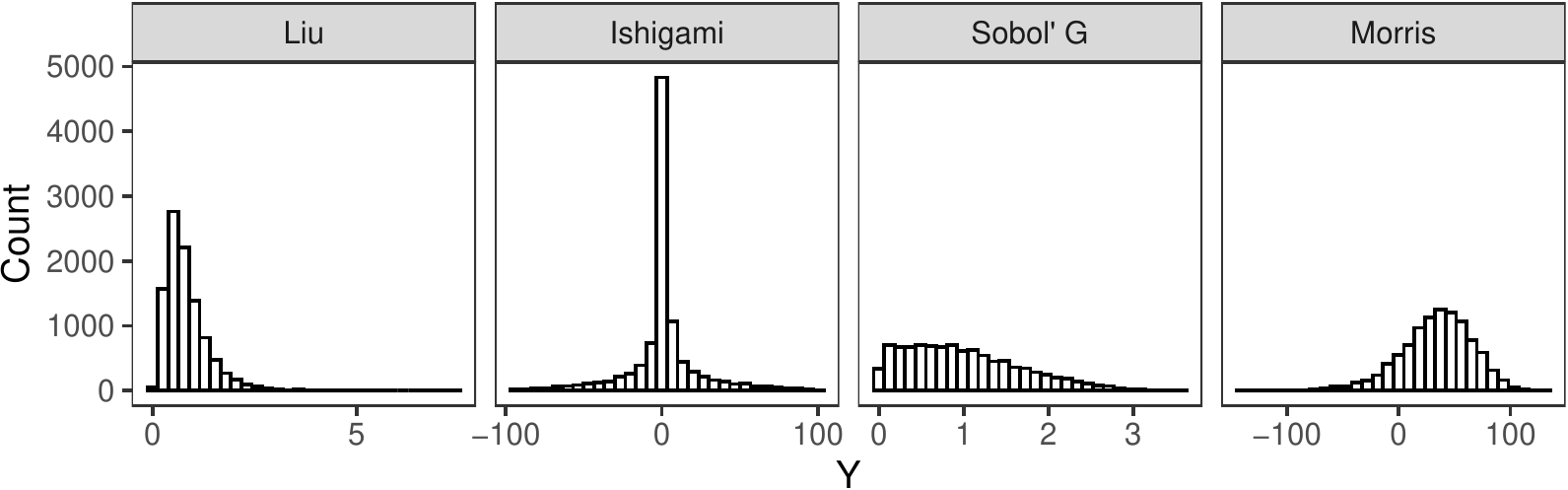}
\caption{Model output of the \textcite{Liu2006a}, \textcite{Ishigami1990}, Sobol' G \parencite{Sobol1998} and Morris \parencite{Morris1991} functions. $N=4000$.}
\label{fig:distributions}
\end{figure*}

For PAWN, we focused on four uncertain parameters: the total number of model runs ($N$), the number of conditioning intervals ($n$), the randomness derived from the sampling of the unconditional model output $Y_U$ ($\varepsilon$), and the summary statistic ($\theta$). For $S_{Ti}^*$, we focused on two: $N$ and $\theta$, the latter reflecting the different estimators existing to compute the total-order effect \parencite{Jansen1999, Homma1996, Sobol2007d} (Table~\ref{tab:parameters}). We assessed how different combinations of values for these uncertain parameters condition the PAWN/$S_{Ti}^*$ index using four different test functions that yield a skewed model output (Fig.~\ref{fig:distributions}): the~\textcite{Liu2006a}'s, which reads as 

\begin{equation}
Y = X_1 / X_2
\label{eq:liu}
\end{equation}

where $X_1\sim \chi^2(10)$ and $X_2\sim \chi^2(13.978)$; the \textcite{Ishigami1990}'s, which reads as

\begin{equation}
Y=\sin(X_1) +  a \sin(X_2) ^ 2 + b X_3 ^4 \sin(X_1)
\label{eq:ishigami}
\end{equation}

where $a=2, b=1$ and $(X_1,X_2,X_3)\sim\mathcal{U}(-\pi, +\pi)$; the Sobol' G \parencite{Sobol1998}'s, which reads as 

\begin{equation}
Y=\prod_{i=1}^{k} \frac{|4 X_i - 2| + a_i}{1 + a_i}
\label{eq:Sobol' G}
\end{equation}

where $k=8$, $X_i\sim\mathcal{U}(0,1)$ and $a=(0, 1, 4.5, 9, 99, 99, 99, 99)$; and the \textcite{Morris1991} function, which reads as

\begin{equation}
\begin{aligned}
Y= & \beta_0 + \sum_{i=1}^{20}\beta_i w_i + \sum_{i<j}^{20}\beta_{i,j} w_i w_j \\
& + \sum_{i<j<l}^{20}\beta_{i,j,l} w_i w_j w_l \\
& + \sum_{i<j<l<s}^{20}\beta_{i,j,l,s} w_i w_j w_l w_s
\end{aligned}
\label{eq:morris}
\end{equation}

where $w_i=2(X_i - 0.5)$ for all $i$ except for $i=4,5,7$, where $w_i=2(1.1X_i / (X_i + 0.1)-0.5)$, $X_i\sim\mathcal{U}(0,1)$, and

\vspace{2mm}
$\beta_i=20, i=1,2,...,10$,

$\beta_{i,j}=-15, i=1,2,...,6$,

$\beta_{i,j,l}=-10, i=1,2,...,5$,

$\beta_{i,j,l,s}=5, i=1,2,...,4$ \parencite{Campolongo2011}.
\vspace{2mm}

\begin{table*}[ht]
\centering
\caption{Summary of the parameters and their distribution for both PAWN ($N, n, \varepsilon, \theta$) and $S_{Ti}^*$ ($N, \theta$). $DU$ stands for discrete uniform.}
\label{tab:parameters}
\begin{tabular}{llc}
\toprule
Parameter & Description & Distribution \\
\midrule
$N$ & Total number of runs & $\mathcal{U}(200,2000)$\\
$n$ & Number of conditioning intervals & $\mathcal{U}(5,20)$\\
$\varepsilon$ & Randomness in the sampling of $Y_U$ & $\mathcal{U}(1,10^3)$\\
$\theta$ & Summary statistic/estimator & $\mathcal{DU}(1,3)$\\
\bottomrule
\end{tabular}
\end{table*}

We chose these four skewed benchmark functions in order to  provide PAWN with a favourable test ground: density-based measures might better characterize the uncertainty in skewed model outputs than variance-based measures as the former do not rely on any statistical moment \cite{Borgonovo2011a}.

We selected the distributions of $(N,n,\varepsilon,\theta)$ based on previous work on PAWN and the $S_{Ti}^*$ index~\parencite{Pianosi2015, Pianosi2018, Saltelli2010a} and some preliminary tests. Firstly, we observed that $N\approx2000$ was sufficient for the PAWN index of many model inputs to be very close to convergence. Sobol' $S_{Ti}^*$, on the other hand, required a larger number of runs (Fig.~S1). This corroborated previous observations on the faster convergence rate of PAWN compared to Sobol' indices \parencite{Pianosi2015, KhorashadiZadeh2017}. By defining $N\sim\mathcal{U}(200,2000)$ we set our study in a scenario where the uncertainty with regards to the required sample size needed to compute robust PAWN indices is moderate. Indeed, this is usually the case: the analyst might have a cap on the total number of model runs available with the computing resources at hand, but no prior information on the minimum sample size required to ensure stable sensitivity indices for all model inputs.

We defined the distribution of $n$ based on~\textcite{Pianosi2018}, who suggested to start with $n=10$ and vary $n$ some units up and down to check its effect (provided that $n>5$). 

For $\varepsilon$, we set $10^3$ different starting points (seeds) for the pseudo-random number sequence used to generate the indices (from 1 to $N$) to sample $Y_U$. This ensured 1) negligible chances of the same seed overlapping with the same value for $N$, thus introducing determinism into a process that should be mainly stochastic, and 2) that the randomness in the sampling of $Y_U$ is assessed in terms of its relative influence in the computation of PAWN. 

With regards to $\theta$, for PAWN we used the mean, the median and the maximum as a summary statistic for the KS values when $\theta=1, \theta=2$ and $\theta=3$. For $S_{Ti}^*$, we used the estimators by \textcite{Jansen1999}, \textcite{Homma1996} and \textcite{Sobol2007d} when $\theta=1, \theta=2$ and $\theta=3$ respectively.

In order to estimate the uncertainty propagated by ($N, n, \varepsilon, \theta$) (resp. $N, \theta$) to PAWN (resp. $S_{Ti}^*$), we created a $(2^{13}, 2k)$ sample matrix for each function using Sobol' quasi-random number sequences, where $k=4$ (resp. $k=2$), and transformed the columns into their appropriate distributions (Table~\ref{tab:parameters}). The first $k$ matrix was labelled $\textbf{A}$ and the second $k$ matrix, $\textbf{B}$. Our model ran row-wise in both the $\textbf{A}$ and $\textbf{B}$ matrices, as follows: based on the information contained in the $v$-th row, it created a Sobol' matrix of size $N^{(v)}$ for PAWN [of size $int(N^{(v)} / (k + 1))$ for $S_{Ti}^*$], and computed either the~\textcite{Liu2006a}, the~\textcite{Ishigami1990}, the Sobol' G \parencite{Sobol1998} or the \textcite{Morris1991} function. Then, for each model input $i$, it either calculated the PAWN$^{(v)}$ index following the conditions set by $n^{(v)}$, $\varepsilon^{(v)}$ and $\theta^{(v)}$, or the $S_{Ti}^{*(v)}$ index according to $\theta^{(v)}$.

We estimated how sensitive PAWN ($S_{Ti}^*$) indices are to uncertainty in ($N, n, \varepsilon, \theta$) (resp. $N, \theta$) by means of Sobol' indices~\parencite{Sobol1993}. For a model of the form $Y=f(X_1, X_2,...,X_k)$, where $Y$ is a scalar and $X_1,X_2,...,X_k$ are independent parameters described by known probability distributions, we can measure how sensible $Y$ is to a given parameter $X_i$ with 

\begin{equation}
V_i=V_{X_{i}}\big[E_{\textbf{X}_{\sim i}}(Y | X_i)\big]
\label{eq:Ex_i}
\end{equation}

where $E_{\textbf{X}_{\sim i}}(Y | X_i)$ is the expected value of $Y$ calculated over all possible values of all parameters except the $i$-th,  which is kept fixed. By dividing Equation~\ref{eq:Ex_i} by the unconditional model output variance, we obtain the first order sensitivity index for $X_i$, which describes the proportion of variance in the model output caused by $X_i$:

\begin{equation}
S_i=\frac{V_i}{V_Y}
\label{eq:Si}
\end{equation}

We can then decompose the unconditional model output variance $Y$ as the sum of conditional variances up to the $k$-th order:

\begin{equation}
V_Y=\sum_{i=1}^{k}V_i+\sum_{i}\sum_{i<j}V_{ij}+...+V_{1,2,...,k}
\label{eq:decomposition}
\end{equation}

where

\begin{equation}
\begin{aligned}
V_{ij}= & V_{X_{i}, X_{j}}\big[E_{\textbf{X}_{\sim i, j}}(Y | X_i, X_j)\big] \\
& - V_{X_{i}}\big[E_{\textbf{X}_{\sim i}}(Y | X_i)\big] \\ 
& - V_{X_{j}}\big[E_{\textbf{X}_{\sim j}}(Y | X_j)\big]
\end{aligned}
\end{equation}

From this, we can derive the second-order index $S_{ij}$, which explains the proportion of variance due to the interaction between $X_i$ and $X_j$:

\begin{equation}
S_{ij}=\frac{V_{ij}}{V_Y}
\label{eq:Sij}
\end{equation}

and so on until order $k$. However, estimating all terms in Equation~\ref{eq:decomposition} is unattainable when $k$ is large, as they result in $2^k-1$. In this case, we can compute the total order index or $S_{Ti}$, which measures the proportion of variance due to the first-order effect of $X_i$ jointly with its interactions with the other parameters \parencite{Homma1996}:

\begin{equation}
S_{Ti}=\frac{E_{X_{\sim i}}\big[V_{X_{i}}(Y | X_{\sim i})\big]}{V_Y}
\end{equation}

For PAWN, since $k=4$, we computed first ($S_i$), second ($S_{ij}$), third ($S_{ijk}$) and total-order ($S_{Ti}$) Sobol' indices of ($N,n,\varepsilon,\theta$). For $S_{Ti}^*$, since $k=2$, we just computed $S_i$ and $S_{Ti}$. In both settings we used the \textcite{Saltelli2010a} and the \textcite{Jansen1999} estimators to compute $S_i$ and $S_{Ti}$ respectively, as per the established best practices. All the workflow is summarised in Fig.~S2 and the $R$ code to replicate our results is available in \href{https://github.com/arnaldpuy/pawn_uncertainty}{GitHub}.

\section{Results}

\subsection{Uncertainty analysis}
\begin{figure*}[ht]
\centering
\includegraphics[keepaspectratio]{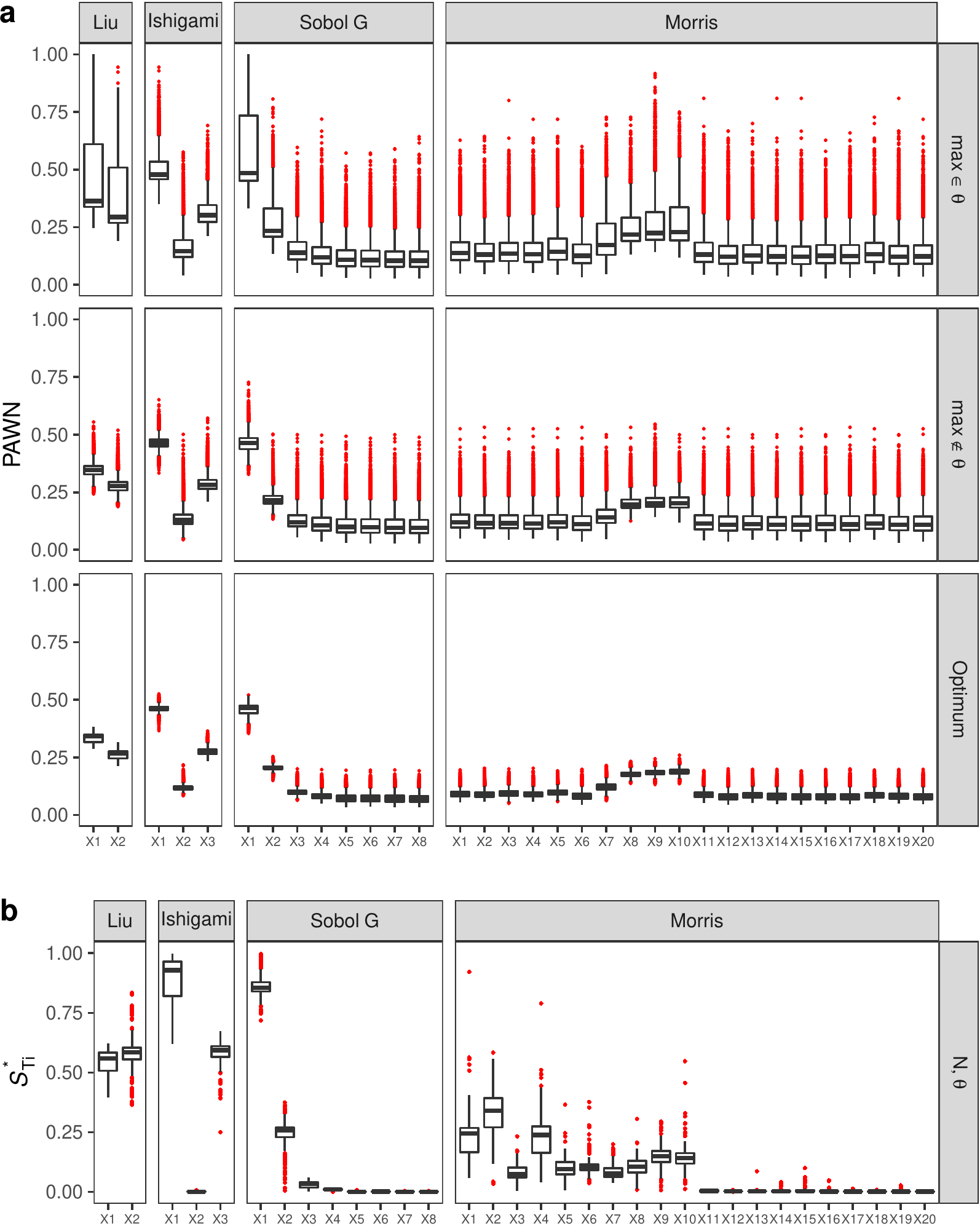}
\caption{Uncertainty in the computation of the sensitivity indices. Outliers are shown in red. a) PAWN. b) $S_{Ti}^*$.}
\label{fig:uncertainty}
\end{figure*}

Fig.~\ref{fig:uncertainty} presents the uncertainty distribution of PAWN and $S_{Ti}^*$ for each model input and function.  The results can be matched against Fig.~S3, where we display how PAWN and Sobol' indices look like once their design parameters are fixed and the total number of model runs is set at $N=4000$.

For PAWN, we ran three simulations (Fig.~\ref{fig:uncertainty}a): 

\begin{enumerate}
\item With $\theta$ including the mean, the median and the maximum as possible summary statistics ($max \in \theta$ setting).

\item With $\theta$ including the mean and the median only ($max \notin \theta$ setting). This aimed at isolating the effect that extreme KS values, which might be obtained for specific conditioning intervals, have in the final PAWN index.

\item With $N \sim \mathcal{U}(2500, 4000)$, $n \sim \mathcal{U}(15,20)$ and ($max \notin \theta$) (\say{Optimum} setting). This latter run reflects an ideal scenario, one in which the number of model runs needed to achieve convergence is known in advance and the uncertainty in $n$ and $\theta$ is reduced to the minimum expression. 
\end{enumerate}

For $S_{Ti}^*$ we run one simulation to assess its sensitivity to the uncertainty in the total number of model runs and the estimator ($N,\theta$) [with $N\sim\mathcal(200, 2000)$] (Fig.~\ref{fig:uncertainty}b). Sobol' indices might eventually take on values outside the range $[0,1]$ due to numerical artifacts created during the computation. This is widely known among sensitivity analysts and managed by considering $(S_{Ti}^*<0)\approx 0$ and $(S_{Ti}^*> 1)\approx 1$. Fig. 2b only presents $S_{Ti}^*\in[0,1]$ to allow for a better comparison with the values produced by the PAWN index. A plot showing the distribution of values with $S_{Ti}^*\notin[0,1]$ is presented in Fig.~S4.

\subsubsection{Factor prioritization}

Fig.~\ref{fig:uncertainty} shows that, in a factor prioritization context, i.e. when the aim is to sort the parameters according to their contribution to the model output variance~\parencite{Saltelli2008}, the uncertainty in the value of the design parameters might cause model inputs to overlap, thus raising the likelihood of producing a biased ranking. In order to get precise figures for this overlap, we computed the coefficient of overlapping, i.e. the area lying under the density curves of two different model inputs, following \textcite{Pastore2018} (see Fig.~S5 for a presentation of Fig~\ref{fig:uncertainty} with density curves instead).

In the case of the \textcite{Ishigami1990} function, the overlap between $X_2$ and $X_3$ is of 10\% if PAWN is used under $max \in \theta$ (4\% under $max \notin \theta$), despite $X_2$ being non-influential. The overlap with $S_{Ti}^*$ is 0. With the \textcite{Sobol1998} G function, PAWN mistakes $X_1$ for $X_2$ and $X_2$ for $X_3$ 11\% and 20\% of the time respectively if $max \in \theta$ (4\% and 10\% if $max \notin \theta$), and might even bias the ranking of $X_2$ and $X_4$ (14\% and 9\% overlap in $max \in \theta$ and $max \notin \theta$ respectively). On the contrary, if $S_{Ti}^*$ is used, the overlap between $X_1-X_2$, $X_2-X_3$ and $X_2-X_4$ is $\sim$0\%, 2\% and 1\% respectively.

The case of the \textcite{Liu2006a} function deserves a specific comment. Fig.~\ref{fig:uncertainty} shows that 29\% of $X_1$ and $X_2$ values overlap if using PAWN under $max \in \theta$ (13\% if $max \notin \theta$), whereas for $S_{Ti}^*$ the degree of overlap is 55\%. \textcite{Liu2006a}, however, stated that $X_1$ was more influential than $X_2$ based on relative-entropy sensitivity methods. \textcite{Pianosi2015} used the \textcite{Liu2006a} function to back up their claim of PAWN outperforming Sobol' indices due to the former being able to discriminate the higher influence of $X_1$ much better than $S_{Ti}^*$. However, the analytical values of $S_{Ti}^*$ in the \textcite{Liu2006a} function are identical at $X_1=X_2=0.546$ (see Table 2 in \textcite{Liu2006a}). This means that the overlap between $X_1$ and $X_2$ shown in Fig.~\ref{fig:uncertainty}b is to be expected and does not result from the sensitivity of $S_{Ti}^*$ to the uncertainties in its own structural design.

\subsubsection{Factor screening}

In a factor screening context, i.e. when the aim is to distinguish influential from non-influential parameters~\parencite{Saltelli2008}, the uncertainty in the PAWN design parameters might also lead to erroneous results, regardless of whether $max \in \theta$ or $max \notin \theta$. In the case of the \textcite{Sobol1998} G function, 10-14\% (7-8\%) of the probability density of $X_2$, which has a non-nihil effect, overlaps with $X_4,...,X_8$, which have no effect at all. The overlap between influential and non-influential model inputs if $S_{Ti}^*$ is used, on the other side, ranges between $\sim$0-2\%. 

The poor screening power of PAWN stands out in the case of the \textcite{Morris1991} function:  18-25\% (1-8\%) of the probability density of $X_8,...,X_{10}$, which have a moderate first-order effect, overlap with that of $X_{11},...,X_{20}$, which are non-influential (Fig. S2A). In the case of $S_{Ti}^*$, the overlap ranges between 0-5\%. The chances of PAWN mistaking relevant for non-relevant parameters is even higher in the case of parameters whose influence in the model output is through interactions only: this is the case of $X_1,...,X_6$ in the \textcite{Morris1991} function, whose degree of overlap with $X_{11},...,X_{20}$, which are non-influential, range between 75-95\%  (Figs.~\ref{fig:uncertainty}--S3). In all these cases, the uncertainty in the design parameters of $S_{Ti}^*$ leads to a 0-4\% overlap between influential and non-influential parameters. 

Fig.~\ref{fig:uncertainty} also shows that in the \say{Optimum} setting the overlap between the model inputs is considerably reduced for PAWN. In the case of the \textcite{Liu2006a} and the \textcite{Ishigami1990} functions, the percentage of overlap goes down to zero. However, the chances of wrongly ranking/screening the model inputs remain non-negligible for both the \textcite{Sobol1998} G and the \textcite{Morris1991} functions. In the former, there is 17--28\% overlap between the slightly influential model input $X_3$ and $X_4,...,X_8$, whose effect can not be differentiated from the approximation error. With regards to the  \textcite{Morris1991} function, the chances of characterizing as non-influential parameters that have a significant non-additive effect in the model output remain very high: the overlap of $X_1,...,X_6$  with $X_{11},...,X_{20}$, for instance, range between 35--90\% (Figs. \ref{fig:uncertainty}). As shown in Fig~S3, the volatility in the computation of PAWN does not allow to distinguish $X_{11},...,X_{20}$ from a dummy, non-influential model input.

\subsection{Sensitivity analysis}
\begin{figure*}[ht]
\centering
\includegraphics[keepaspectratio]{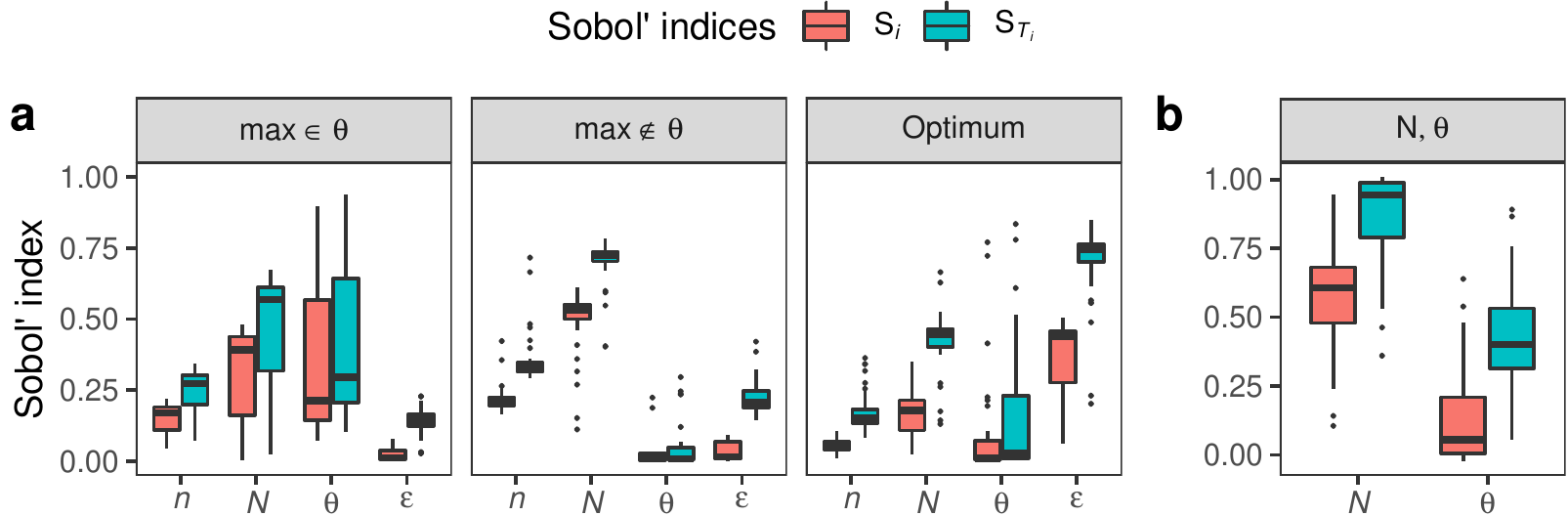}
\caption{Sobol' indices. a) PAWN. b) $S_{Ti}^*$.}
\label{fig:aggregated}
\end{figure*}

Figs.~\ref{fig:aggregated} presents the Sobol' first ($S_i$) and total ($S_{Ti}$) indices for each of the settings of our analysis after pooling the values from all functions and parameters [the Sobol' indices for each function and design parameter are shown in Figs.~S6--S7 (PAWN), and Figs.~S8--S9 ($S_{Ti}^*$)]. Figs.~\ref{fig:aggregated} thus inform on how much each design parameter contribute uncertainty to each sensitivity index. In Fig.~S10 of the Supplementary Information file, we prove that the results displayed in Fig.~\ref{fig:aggregated} are robust without having to offset the stronger weight that the Morris function might have in defining the trends due to its much larger number of model inputs.

As shown in Fig.~\ref{fig:aggregated}a, the first-order effect of $\theta$ and $N$ is much more variable in the $max \in \theta$ setting, suggesting that their degree of contribution to the PAWN index uncertainty might considerably be function-dependent. This variability is highly reduced in the $max \notin \theta$ setting, which provides a more robust account of the extent to which each design parameter contributes to define the PAWN index. In this setting, the selection of the initial sample size ($N$) and the number of conditioning intervals ($n$) convey up to c. 60\% and c. 20\% of the PAWN index uncertainty respectively. The stochasticity in the sampling of $Y_U$ ($\varepsilon$) is mostly influential through interactions in both settings, whereas the selection of the summary statistic ($\theta$) has a nearly nihil effect in the $max \notin \theta$ setting. Remarkably, interactions are also significant in the \say{Optimum} setting, especially those involving $\varepsilon$, $n$ and $N$.

\begin{figure*}[ht]
\centering
\includegraphics[keepaspectratio]{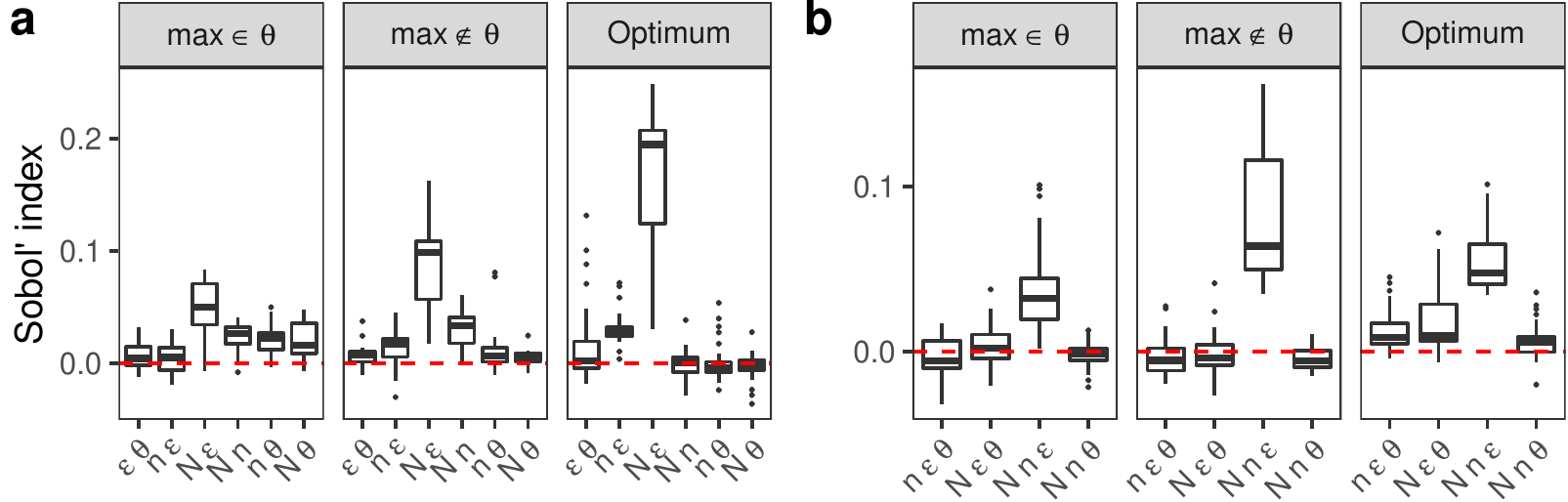}
\caption{High-order interactions between the PAWN design parameters. Only those boxplots with all values above the red dashed horizontal line reflect a true effect. a) Second-order interactions. b) Third-order interactions.}
\label{fig:second_third}
\end{figure*}

To gain further insights into the structure of the non-additivities in PAWN, we computed second and third-order effects, shown in Fig.~\ref{fig:second_third} (the second and third-order Sobol' indices for each function and parameter are shown in Figs.~S11--S14). In the $\max \notin \theta$ setting, the interactions that have a significant effect on the model output involve the initial sample size ($N$) with the number of conditioning intervals ($n$) or the stochasticity in the sampling of $Y_U$ ($\varepsilon$). Such second-order effects might contribute up to 15\% of the PAWN index uncertainty. This also applies to the \say{Optimum} setting, where the interaction between $N$ and $\varepsilon$ or $n$ and $\varepsilon$ has an even higher effect (up to 25\% uncertainty). These three design parameters have significant third-order interactions in all three settings (Fig.~\ref{fig:second_third}b).

Regarding $S_{Ti}^*$ (Fig.~\ref{fig:aggregated}b), both $N$ and $\theta$ have a non-negligible first-order effect on the index. The wide boxplots suggest that both the extent of their influence and the degree of non-additivities in defining the final $S_{Ti}^*$ index depends on the model under scrutiny (Figs.~S8--S9). 

\section{Discussion and conclusions}

Sensitivity analysis is an important tool to check the robustness of a model in the context of its validation. However, also the measurement of the sensitivity of the output variables to input parameters rests on modelling hypotheses, i.e. sensitivity analysis is a modelling process per se, based on the use of an algorithm -- like the models being investigated. Here we tested the robustness of this aspect of the modelling process by assessing the dependency of the PAWN index to its design parameters, and matched the results against the Sobol' total-order index ($S_{Ti}^*$). Our work thus sheds some light on the uncertainties concealed in the last stage of any sensitivity analysis: the implementation of the sensitivity algorithm. This practice could help ensuring the adequacy and the range of applicability of the tools which modellers deploy to improve the quality of their modelling exercises. 

Two elements emerge from our work: the PAWN index is more sensitive to the design parameters than $S_{Ti}^*$, and this sensitivity has a complex pattern which makes the use of PAWN, or better the tuning of the PAWN design parameters, a delicate task. The chances of incurring in false positives (mistaking a non-influential parameter for influential, with a waste of computational resources) and/or false negatives (mistaking an influential parameter for non-influential, with a loss of useful information) are significant even when the design parameter space does not include the maximum as a possible summary statistic, but only measures of central tendency. Even in an ideal setting, where the uncertainty in $n$ and $\theta$ is highly reduced and $N$ ensures convergence, the PAWN index might be incapable to differentiate between non-influential model inputs and influential model inputs whose effect in the model output is fully through interactions. \textcite{Mora2019} also observed that the high level of noise in PAWN might produce biased rankings when the model inputs have different orders of magnitude in their contribution to the response.

PAWN especially underperformed when used in the Sobol' G \parencite{Sobol1998} ($k=8$) and the \textcite{Morris1991} function ($k=20$). This raises a red flag for analysts when using PAWN in high-dimensional models such as those commonly being employed in the Earth and Environmental Sciences domain, which might easily include tens of parameters \parencite{Sheikholeslami2019a}. In these contexts, the use of PAWN might indeed allow to significantly reduce the number of model runs required to conduct a sensitivity analysis, but at the expense of significant risks of obtaining a biased result due to its structural design.

The fact that the sensitivity of PAWN to its design parameters is complex, including important interactions up to the third order effect, implies that finding the perfect range of design parameters to use PAWN safely and efficiently is not easy. The significant non-additivity of PAWN only unfolds once the values of its main design parameters are moved simultaneously within reasonable uncertainty ranges; this is, when all the forking paths and divergences leading towards its computation are assessed at once. Instead, in the paper where the \say{generic approach} is presented, \textcite{Pianosi2018} analysed the influence of ($N,n,\theta$) on PAWN by combining different discrete point-estimates for $N$ and $n$ (see their Fig. 3), or by changing the value for either $N$ or $n$ while keeping the other design parameters fixed (see their Figs. 6 and 7). This approach is very similar to a one-at-a-time (OAT) sensitivity analysis, a method that may fail to detect interactions between model parameters due to its incomplete examination of the uncertainty space \parencite{Saltelli2010b}. Our work shows that even by increasing $N$, ensuring a \say{high} $N/n$ ratio or fixing $\theta$ at a central tendency measure, significant interactions between $Nn$, $N\varepsilon$ or between $Nn\varepsilon$ make PAWN considerably difficult to tame.

In our study we also observed the existence of potential interactions between the total number of model runs $N$ and the estimator $\theta$ for $S_{Ti}^*$. However, the variance in $S_{Ti}^*$ deriving from the uncertainty in its design parameters is comparatively much smaller, thus reducing the chances of obtaining a biased ranking or producing a wrong screening. It should be stated that our research design for $S_{Ti}^*$ included the use of different estimators out of our willingness to fully explore its potential uncertainty space. However, there is actually no uncertainty in $\theta$ when computing Sobol' total-order indices: \textcite{Saltelli2010a} showed that the best estimate of the degree to which a given model input interacts with the rest is obtained with the \textcite{Jansen1999} estimator, which should be the default choice. This means that the uncertainty in the design parameters of $S_{Ti}^*$ actually narrows down to the total number of model runs $N$ only. It is thus to be expected that given two algorithms, the one depending on a higher number of design parameters will be the more delicate to use. A similar discussion led \textcite{Campolongo2011} to conclude that the screening method of Morris could be effectively substituted by $S_{Ti}^*$. 

We would like to stress that there are many other uncertainties beyond those embedded in sensitivity indices that have the potential to condition the results of a given sensitivity analysis. These do not only involve quantifiable elements such as the selection of the distributions, sample size or the model output form \cite{Kaplan1981, Aven2020, Noacco2019}, but also uncertainties in the entire knowledge and model-generating processes \cite{Saltelli2013}. Algorithm uncertainties ingrained in the design parameters of sensitivity indices, however, are especially important for they condition the computational path towards the final sensitivity value. Although several alternative routes might be available, these multiple options are eliminated once the computation is triggered, yielding the illusion of a deterministic process.  We believe that a good way to know whether a sensitivity index is robust is to check how volatile it is when all paths involved in its effective computation are walked at once.

The present findings do not suggest discarding PAWN as a sensitivity measure. Moment independent measures have a role to play in sensitivity analysis of output with long-tailed distributions. Additionally, they may find an ideal use in settings where the output of interest is itself in the form of a difference between two cumulative distributions. The present analysis aims at encouraging developers of sensitivity indices to fully explore the structural uncertainty of their algorithms in order to deliver transparent and robust sensitivity tools.

\section*{Acknowledgements}
We thank Francesca Pianosi, Razi Sheikholeslami, Thorsten Wagener and two anonymous reviewers for their constructive comments on previous versions of this manuscript. This work has been funded by the European Commission (Marie Sk\l{}odowska-Curie Global Fellowship, grant number 792178 to A.P.). 

\printbibliography

\end{document}